\documentclass[letterpaper, 10 pt, conference]{ieeeconf}

\IEEEoverridecommandlockouts                              % This command is only
\usepackage{booktabs}

\usepackage[dvipsnames]{xcolor}
\usepackage[normalem]{ulem}
\usepackage{xcolor}
\usepackage{cite}
\usepackage{amsmath,amssymb,amsfonts}
\usepackage{algorithmic}
\usepackage{graphicx}
\usepackage{textcomp}
\usepackage{graphics} % for pdf, bitmapped graphics files
\usepackage{epsfig} % for postscript graphics files
\usepackage{times} % assumes new font selection scheme installed
\usepackage{csquotes}
\usepackage{multirow}
\title{\LARGE \bf 
MPC strategies for density profile control with pellet fueling in nuclear fusion tokamaks under uncertainty }
\author{Christopher A. Orrico*, Hari Prasad Varadarajan*, Matthijs van Berkel, Lennard Ceelen,\\ Thomas O. S. J. Bosman, W. P. M. H. Heemels, Dinesh Krishnamoorthy%
\thanks{This publication is part of the project \textit{Balls to the Wall} (project no. 19695) of the research programme NWO Talent Programme VIDI, financed in part by the Dutch Research Council (NWO). DIFFER is an institute of the NWO.}
\thanks{This work has been carried out within the framework of the EUROfusion Consortium, funded by the European Union via the Euratom Research and Training Programme (Grant Agreement No. 101052200—EUROfusion). Views and opinions expressed are however those of the author(s) only and do not necessarily reflect those of the European Union or the European Commission. Neither the European Union nor the European Commission can be held responsible for them.}
\thanks{We thank Florian Koechl, Peter Fox and the JINTRAC/HFPS development team for their support with high-fidelity plasma simulations. }
\thanks{*These authors contributed equally to this work.}
\thanks{C. A. Orrico, D. Krishnamoorthy, and W. P. M. H. Heemels are with the Dept. of Mechanical Engineering, Eindhoven University of Technology, 5600 MB Eindhoven, The Netherlands (e-mail: c.a.orrico@tue.nl, d.krishnamoorthy@tue.nl, w.p.m.h.heemels@tue.nl).}
\thanks{H.P. Varadarajan, M. van Berkel, L. Ceelen, and T. O. S. J. Bosman are with DIFFER - Dutch Institute for Fundamental Energy Research, 5612 AJ Eindhoven, The Netherlands (e-mail: h.varadarajan@differ.nl, m.vanberkel@differ.nl, l.ceelen@differ.nl, t.o.s.j.bosman@differ.nl).}
}

\begin{document}

\maketitle
\thispagestyle{empty}
\pagestyle{empty}

\begin{abstract}
%Control of the density profile with pellet fueling for the ITER nuclear fusion tokamak involves a multi-rate nonlinear system with safety-critical constraints, input delays, and discrete actuators with parametric uncertainty. We propose a mixed interger model predictive (MI-MPC) control strategy for this case study, wherein we follow a three-pronged approach to manage computational complexity and uncertainty. First, we use dynamic mode decomposition with control to produce a reduced-order control-oriented model from synthetic system identification data taken from high-fidelity simulations of the ITER plasma. Next, we devise a multi-stage scenario MI-MPC (msMI-MPC) strategy using scenario trees, utilizing principal component analysis to reduce the number of scenario trees needed to capture the parametric uncertainty. Finally, we introduce our novel multi-stage scenario penalty term homotopy MPC (msPTH-MPC) to reduce the computational burden of msMI-MPC. We compare the performance and safety of the msMI-MPC and msPTH-MPC strategies against a nominal MI-MPC in plant simulations, demonstrating the first predictive density control strategy with uncertainty handling, viable for real-time pellet fueling in ITER.

Control of the density profile based on pellet fueling for the ITER nuclear fusion tokamak involves a multi-rate nonlinear system with safety-critical constraints, input delays, and discrete actuators with parametric uncertainty. To address this challenging problem, we propose a multi-stage MPC (msMPC) approach to handle uncertainty in the presence of mixed-integer inputs. While the scenario tree of msMPC accounts for uncertainty, it also adds complexity to an already computationally intensive mixed-integer MPC (MI-MPC) problem. To achieve real-time density profile controller with discrete pellets \textit{and} uncertainty handling, we systematically reduce the problem complexity by (1) reducing the identified prediction model size through dynamic mode decomposition with control, (2) applying principal component analysis to reduce the number of scenarios needed to capture the parametric uncertainty in msMPC, and (3) utilizing the penalty term homotopy for MPC (PTH-MPC) algorithm to reduce the computational burden caused by the presence of mixed-integer inputs. We compare the performance and safety of the msMPC strategy against a nominal MI-MPC in plant simulations, demonstrating the first predictive density control strategy with uncertainty handling, viable for real-time pellet fueling in ITER.

\end{abstract}

%\begin{IEEEkeywords}
%Emerging control applications, Energy systems, Hybrid systems, Predictive control for nonlinear systems, Uncertain systems
%\end{IEEEkeywords}

\section{Introduction}
\begin{figure}[t!]
    \centering
    \includegraphics[width=\linewidth]{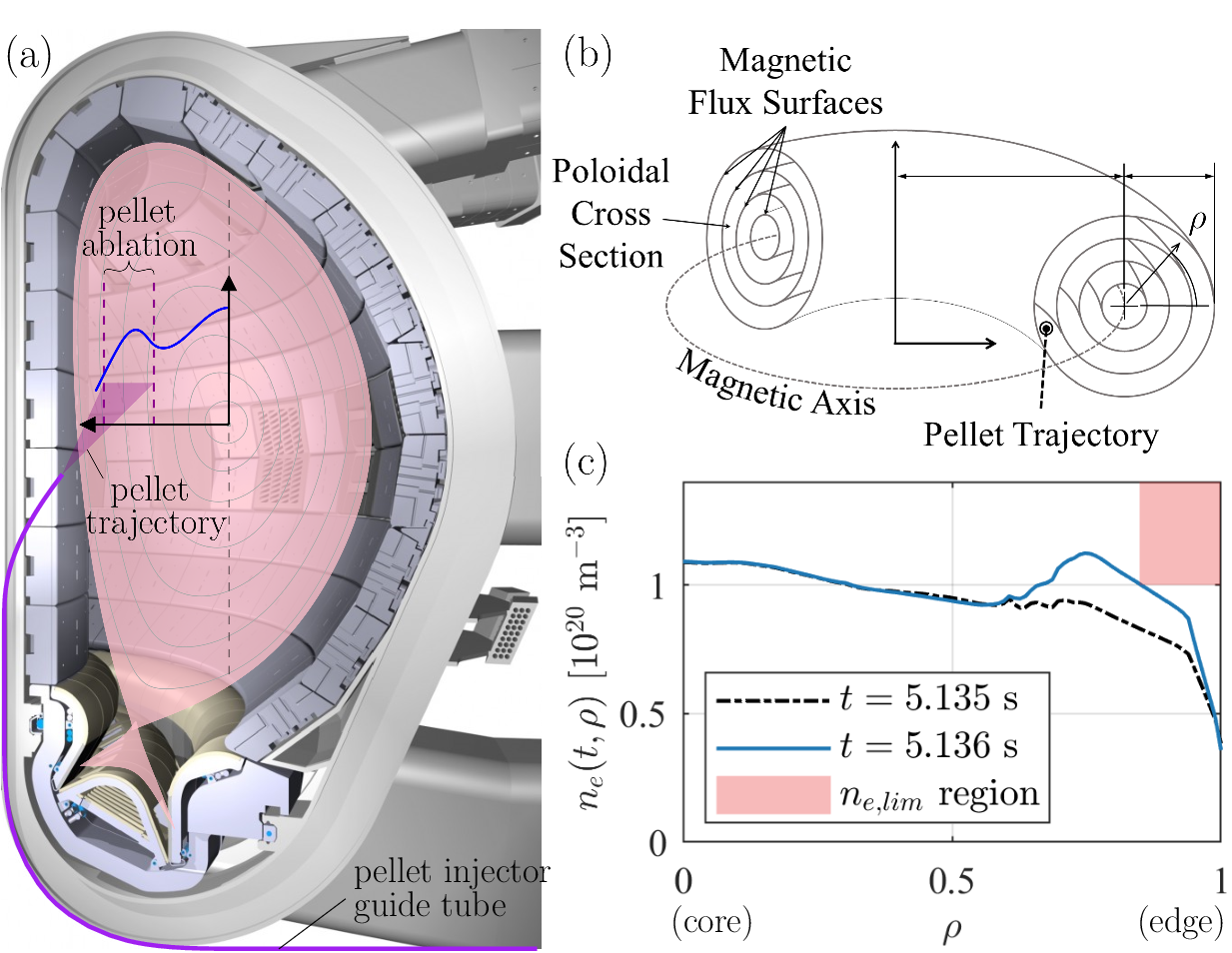} 
    \caption{(a) Diagram of a poloidal cross-section of the ITER tokamak with the plasma region in pink. Image courtesy of the ITER organization with modifications \cite{ITER_VV}. The trajectory of the pellet (purple) into the plasma core is overlaid on a sketch of the 1D $n_{e}(\rho)$ profile (blue). $n_{e}(\rho)$ is constant on each \textit{magnetic flux surface}, which are denoted in grey in (a) and labeled in (b). (b) The 3D plasma, sketched as a cylindrical approximation labeled with the flux label $\rho$, where $\rho = 0$ at the plasma center and $\rho = 1$ at the plasma edge. (c) Realization of the 1D $n_e(t,\rho)$ profile immediately before ($t = 5.135$ s) and after ($t = 5.136$ s) a pellet enters the plasma (c.f Fig. \ref{fig:simulation}). The region above $n_{e,lim}$ that $n_{e}(t,\rho)$ cannot enter is shown in red.}
    \label{fig:iterdiagram}
\vspace{-0.5cm}
\end{figure}

Nuclear fusion is expected to play a crucial role in meeting the world's carbon-free energy demands over the next century. Tokamak fusion reactors, the most technologically developed method for nuclear fusion, require real-time density control (along with control of heating, current drive, etc.) of the deuterium-tritium (D-T) plasma of density $\sim 10^{20}$ particles$~\cdot~ m^{-3}$ at $15$ keV ($\sim 100$ million $^{\circ}$K) to sustain fusion power production \cite{garzotti2019,bosman2023, orrico2023,pajares2017}. In the next generation of fusion tokamaks, such as ITER, high-velocity frozen fuel pellets of D$_2$ and T$_2$ will be injected into the plasma such that the electron density profile $n_e(t,\rho)$ approaches a desired reference $n_e^{ref}(t,\rho)$ \cite{Baylor2016,garzotti2019}. Hence, a pellet fueling controller is needed to regulate the \textit{density profile}, the behavior of which can be modelled as a nonlinear $1$D partial differential equation (PDE) approximating (c.f. Fig \ref{fig:iterdiagram}) volumetric particle transport from the core to the edge of a tokamak plasma. 

Pellet fueling creates several challenges for density profile control. The pellet time-of-flight ($t_{TOF}$) between the moment a pellet is fired and when it enters the plasma, combined with the time necessary for pellet particles to drift to the plasma core, creates significant actuator delay. Moreover, we are dealing with safety-critical processes, as we must \textit{never} exceed the safety-critical plasma edge density limit $n_{e,lim}$ \cite{Giacomin2022}. Exceeding $n_{e,lim}$ leads to a plasma disruption, which would cause significant, possibly irreparable damage to the ITER tokamak. To solve this control problem, model predictive control (MPC) serves as a natural candidate as it can incorporate actuator delay in the prediction model and account for state constraints.

However, pellet injectors are discrete actuators, with each pellet's ablation causing a large \textit{jump} in the plasma density. MPC for such systems with discrete actuation results in a mixed integer MPC (MI-MPC) that requires solving mixed-integer programs (MIPs). MIPs are known to be $\mathcal{NP}-$hard (i.e. are difficult to solve in polynomial time) \cite{Bemporad1999}. Complicating matters further, turbulent transport, which dictates the evolution of the plasma density profile, is nonlinear. Turbulence-driven nonlinearities are especially pronounced in the pellet ablation and deposition profiles, leading to uncertain actuator dynamics \cite{angioni2017density, panadero2023}. The performance and safety of nominal linear MPC degrades significantly under such uncertainties. Consequently, any successful MPC strategy should consider uncertainty to prevent possible violation of the edge density limit\cite{bosman2023, orrico2025}. In spite of this, uncertainty has been neglected in nuclear fusion literature \cite{pajares2017,bosman2023,orrico2023,orrico2025} until this work.
%However, pellets are discrete actuators, as each pellet causes a large and sudden jump in the plasma density. MPC for such systems with discrete actuation requires solving mixed-integer programs (MIPs), which are known to be $\mathcal{NP}-$hard (i.e. are difficult to solve in polynomial time) \cite{Bemporad1999}. Complicating matters further, turbulent transport, which dictates the evolution of the plasma density profile, is highly nonlinear. Turbulence-driven nonlinearities are especially pronounced in the pellet ablation and deposition profiles, leading to uncertain actuator dynamics \cite{angioni2017density, panadero2023}. The performance and safety of MPC-based methods degrades significantly under such uncertainties which the MPC strategy must consider to prevent possible violation of the edge density limit\cite{bosman2023, orrico2025}. Despite this, uncertainty has been largely neglected in the fusion control literature until this work (CHRIS MOVE CITATIONS FROM MAIN CONTRIBUTUIONS). 

Yet, many robust and stochastic methods exist for addressing the challenge of uncertainty in MPC \cite{Mayne2016}. Stochastic MPC minimizes the average performance typically with chance constraints and hence is not suited for problems with safety critical constraints. Classical robust MPC, on the other hand, results in a min-max formulation, which is generally intractable for nonlinear problems. Although the problem is tractable for linear systems with quadratic costs and convex constraint sets, optimizing over open loop control sequences results in an overly-conservative solution and optimizing over feedback policies results in an infinite dimensional problem \cite{Mayne2016}. In the case of binary decision variables, the constraint set is no longer convex, making it unsuitable for our problem. An alternative approach to robust MPC is the so-called tube MPC that optimizes a nominal control trajectory while accounting for unknown-but-bounded disturbances. For both linear and nonlinear tube MPC, the most widely applied approach is to separate the uncertain dynamics from the nominal state trajectory through use of an ancillary controller in addition to the nominal controller. The nominal controller then regulates the state trajectory, while the ancillary controller ensures the state remains within a robust invariant \textit{tube} around the nominal trajectory. While many approaches exist for constructing ancillary controllers for linear and nonlinear robust MPC \cite{langson2004, Mayne2007, Mayne2016 ,singh2016,Sasfi2023}, they generally require input constraints to be tightened in one way or another. Input constraint-tightening on discrete actuators is not feasible, meaning robust MPC methods with ancillary controllers are ill-suited for density profile control with discrete pellets.  

%

%Robust tube-based MPC (T-MPC) \cite{langson2004} controllers are widely used to handle parametric uncertainty. In T-MPC, a continuous controller (used in conjunction with a nominal MPC controller) drives the state to a pre-constructed robust control invariant (RCI) set, ensuring the system trajectory stays within desired bounds despite the presence of an unknown-but-bounded disturbance. However, without continuous inputs, computing the RCI set becomes difficult or even impossible. Additionally, the presence of integer decision variables renders the problem non-convex, violating the convexity assumption that underpins T-MPC. Consequently, the T-MPC approach ill-suited for density profile control with discrete pellets.

Another approach for handling uncertainty is through a method called multi-stage scenario MPC (msMPC), which is a tractable approximation of optimization over feedback policies \cite{scokaert1998scenarioMPC, lucia2013}. The idea of msMPC is to optimize several discrete control trajectories, one for each disturbance realization (called a scenario) in a finite set, that capture the impact of disturbances. The control trajectories are optimized over multiple scenario branches, creating a \textit{scenario tree}, where each branch assumes a certain disturbance realization \cite{lucia2013}. Critically, the msMPC approach can be extended to systems with discrete actuators in theory, and hence is the most suitable MPC under uncertainty formulation for systems with discrete actuators. 

While msMPC provides an avenue to address uncertainty in predictive density profile control with pellet fueling for the ITER tokamak, it only further increases the computational burden of the already $\mathcal{NP}-$hard MI-MPC problem. Hence, to handle uncertainty in real time, we must reduce computational complexity wherever possible. 

\vspace{0.5em}
\noindent \textit{Main Contributions:} The multistage mixed-integer MPC (msMI-MPC) strategy developed in this work is the first predictive density profile controller for tokamaks, (referred to hereon as density control) which simultaneously accounts for uncertainty in pellet deposition and treats pellet injection as integer decisions. For this controller to also be real-time capable, we present a three-pronged approach for reducing computational complexity: 

\begin{enumerate}
    \item We utilize existing system identification data of our system to perform model order reduction using dynamic mode decomposition with control (DMDc) \cite{proctor2016dynamic}. 
    \item To reduce the number of scenario tree branches of msMPC, we use a data-driven scenario selection procedure based on principal component analysis (PCA) \cite{krishnamoorthy2018data}. Applying this method on the set of parameter realizations measured \textit{a priori} from the system identification data, we reduce the set of parameter realizations needed to capture parametric uncertainty.
    \item To handle the computational complexity of MIPs, we reformulate msMI-MPC using the modified penalty term homotopy algorithm for MPC (PTH-MPC), which relaxes the MIP to a series of continuous optimization problems, thereby enabling real-time computation.
\end{enumerate}

We compare the performance and safety of two resulting msMPC strategies against nominal MI-MPC in plant simulations of density control.

\vspace{0.5em}
%\paragraph*{Main Contributions} \hspace{-0.2em} We compare the performance and safety of two resulting msMPC strategies against nominal mixed-integer MPC in plant simulations of density control, demonstrating the first steps towards a MPC solution with uncertainty handling viable for real-time pellet fueling in the ITER tokamak. In contrast to previous works, this work explicitly accounts for uncertainty in the deposition profiles of discrete pellets for density control in tokamaks. 
%% UPDATE ME FOR THIS PAPER
\noindent \textit{Notation:} $\mathbb{R}^n$ denotes the vector space of $n-$dimensional of real vectors. For a vector $a$ and square matrix $B$, the notation $\|a\|_B^2 = a^\top B a$. We define $A^\dagger$ as the pseudoinverse of matrix $A$. We denote $s \in _R \mathbf{S}$ as the random selection of realization $s$ from set $\mathbf{S}$. Discrete time is given as $t \in \mathbb{N}^+$ ms (uniform sampling $\tau_s = 1$ ms). From time $t$, the predicted time is taken as $k \in \mathbb{N}^+$ ms discrete time steps from $t$, also with sampling $\tau_s$. The dimensionless flux label $\rho \in [0,1]$ defines the distance from the plasma magnetic axis in $1$D (c.f. Fig \ref{fig:iterdiagram}, \cite{Blanken2018} and the references therein). $n_e(t,\rho)$ denotes the electron density in $10^{20}$ m$^{-3}$ at time $t$ and flux label $\rho$. The average core density $\bar{n}_{e,core}(t)$ is defined as the volume-averaged density of $n_e(t,\rho)$ over $\rho \in [0.0,0.4]$ at time $t$ \cite{Lang_2022}. The edge density $n_{e,edge}(t)$ is defined as the density $n_e(t,\rho)$ at $\rho = 0.85$ for time $t$ \cite{Giacomin2022}. The edge density limit, denoted $n_{e,lim}$ and evaluated at $n_{e,edge}(t)$ is taken as $1 \times 10^{20}$ m$^{-3}$ for the ITER plasma scenario explored in this work \cite{Giacomin2022}.

\section{Reduced Order Control-Oriented Modeling}
\label{subsec:ControlProblem}

\noindent \textit{Problem Setting:} For the density control problem, we seek to track a reference $n_e^{ref}(t,\rho)$ over $\rho \in [0.0,0.4]$ using pellet fueling without exceeding $n_{e,lim}$ (taken as a hard constraint $n_{e,edge}(t) \leq n_{e,lim}$ for all $t\geq0$). The flux label $\rho$, analogous to a cylinder radius, is the spatial dimension defining the distance from the plasma core to the plasma edge in the nonlinear PDE $\partial n_e(t,\rho)/\partial t$ which governs the evolution of plasma density (for a full description of $\partial n_e(t,\rho)/\partial t$, c.f  \cite{Blanken2018,dnestrovskij2013,felici2018real}). As the ITER tokamak will require effective controllers at the moment it is operational, controller development is now done using JINTRAC, a high-fidelity simulator for ITER tokamak plasma discharges \cite{romanelli2014}. Due to the computational complexity of the plasma turbulence models \cite{staebler2020geometry} necessary to compute particle transport in JINTRAC (ranging in wall-clock times of $\sim$10-100 hours per second of simulated plasma), iterating controller designs over several instances of simulation is generally impractical. 

Rather than re-running the JINTRAC simulations for each controller design iteration, we regress a linear-parameter varying (LPV) surrogate plant model using the results of synthetic system identification experiment in a 12s JINTRAC simulation of the ITER 15 T/5.3 MA plasma scenario \cite{romanelli2014, garzotti2019, orrico2025}. The discretization of $\rho$ necessary to capture local density evolution $n_e(t,\rho) \in \mathbb{R}^{n_y}$ is typically $n_y = 100$, which introduces additional computational burden to MPC were we to directly use the PDE $\partial n_e(t,\rho)/\partial t$ as our state prediction model. Instead, we use the DMDc method on our system identification data to compute the least-squares regression of a reduced-order state-space model of state order $n_x = 4$ and output order $n_y = 100$, greatly reducing the model (and therefore computational) complexity. The reduced-order, LPV state-space model with input delay $d$ describing the evolution of $n_e(t,\rho)$, given as, 

\begin{subequations} \label{eq:plantmodel}
\vspace{-1em}
    \begin{align}
    x(t+1) &= Ax(t) + B(p(t))u(t-d) \label{eq:plantstate} \\
    y(t) &= Cx(t) \label{eq:plantoutput}
    \end{align}
\end{subequations}

\noindent captures the uniform spatial discretization of $n_e(t,\rho)$ at discrete time $t$ (c.f. Fig \ref{fig:iterdiagram}c) as the output $y(t) \in \mathbb{R}^{n_y}$ over $\rho \in \mathbb{R}^{n_y}$. The state $x(t) \in \mathbb{R}^{n_x}$, captures the dynamic evolution of $n_e(t,\rho)$ through reduced-order state and input matrices $A \in \mathbb{R}^{n_x \times n_x}$ and $B \in \mathbb{R}^{n_x \times n_u}$. The model reduction in \eqref{eq:plantmodel} represents the physical state projected onto a low dimensional basis formed by a proper orthogonal decomposition of the data snapshot matrix (constructed using the sequence of state and input system identification data) \cite{proctor2016dynamic, brunton2022data}. The input $u(t) \in \{0,1\}$ is constrained to the decision to fire a pellet ($u(t) = 1$) or not ($u(t) = 0$). 

We then compare (1) against a validation JINTRAC simulation of 9.7s duration with similar plasma conditions to the simulation used for identification, evaluating the estimated output in terms of 1-step prediction root-mean square error (RMSE) and the RMSE of an open loop simulation of (1) starting from the same initial state as the validation simulation. The mean 1-step prediction RMSE of (1) is $8.481 \times 10^{-3}$ and the mean RMSE of the open loop simulation is $2.267 \times 10^{-2}$ (note here that $\mathcal{O}(y) =1$). The low RMSE values indicate that the linear reduced order model captures the nonlinear turbulent transport dynamics of JINTRAC well during autonomous state evolution around the expected operating point. However, we see that the 1-step prediction error after a pellet is fired is high, with a mean RMSE of $4.129 \times 10^{-2}$, an order of magnitude higher than the mean prediction error when $u(t) = 0$. This significant uncertainty is due to the nonlinear plasma behavior pellet deposition and ablation, described in detail in \cite{panadero2023}. As it is the largest source of plant-model mismatch and is seen in previous work to result in MPC constraint violation \cite{bosman2023}, we treat it the main focus of uncertainty handling in this work.

We capture the plant-model mismatch resulting from nonlinear pellet dynamics as input-affine disturbances in the linear parameterization of
\vspace{-0.2em}
\begin{equation}
    B(p(t)) = p(t) + B_0 \in \mathbb{R}^{n_x \times n_u},
\end{equation}
\noindent where $p(t) \in \mathbb{R}^{n_x}$ and $B_0 \in \mathbb{R}^{n_x \times n_u}$ is the nominal value of $B(p(t))$ computed with DMDc. From the system identification data, $m$ realizations of $p(t)$ are represented by a data matrix $\mathbf{P} \in \mathbb{R}^{m \times n_x}$. In section \ref{sec:results}, we use \eqref{eq:plantmodel} and random realizations of $p(t) \in_R \mathbf{P}$ as a lightweight LPV plant simulator capturing the pellet-driven nonlinear input dynamics present in the original high-fidelity plasma simulation in order to devise a control strategy with uncertainty handling. 

To ensure the plant is analogous to the ITER plasma control system (PCS) and pellet injection system (PIS) \cite{Baylor2016}, we adopt the following multi-rate and input delay dynamics:
\begin{itemize}
    \item Dynamics: $x(t)$ and $y(t)$ evolve on the timescale of $\tau_s = 1$ ms.
    \item Control updates: We set the rate at which pellets are available in the ITER PIS as $10$ Hz \cite{Baylor2016}. We take a slow control time sample $\tau_c = 100\tau_s$ \cite{orrico2023}. At $t \in \{\tau_c, 2\tau_c, 3\tau_c, \ldots  \}$, a pellet decision is made using MPC where $u(t) \in \{0,1\}$. Between control updates, $u(t) = 0 ~\forall~ t \notin \{\tau_c, 2\tau_c, 3\tau_c, \ldots  \}$. 
    \item Input delay: For real-time implementation, the PIS must wait for a control decision within a computation time $t^{lim}_{cpu} = 100$ ms. Before entering the plasma, the pellet must travel along a guide tube (c.f. Fig \ref{fig:iterdiagram}a), adding the delay $t_{TOF} = 35$ ms \cite{Baylor2016}. While current devices neglect this TOF, it is large enough in ITER that the controller must take it into account. Thus, a pellet $u(t-d)$ fired at $t-d$ will actually enter the plasma at $t$, where $d = t_{cpu}^{lim} + t_{TOF} = 135$ ms. 
\end{itemize}
\noindent For further details on the pellet injection system, c.f. \cite{Baylor2016, bosman2023, orrico2025} and the references therein. 

\section{Multi-Rate, Mixed-Integer MPC Strategies}
\label{sec:MPCapproach}

We study three multi-rate MPC strategies for density control. First, we briefly describe our baseline nominal mixed-integer MPC (MI-MPC) formulation with a multi-rate approach to handle slow control updates and fast dynamics \cite{lee1992model, jiang2011explicit}. Then, we extend it to a multi-stage scenario MI-MPC (msMI-MPC) formulation, which accounts for the parametric uncertainty in $B(p(t))$ using a scenario tree. Finally, we introduce our novel multi-stage scenario penalty term homotopy MPC (msPTH-MPC), which reduces computational complexity to meet our real-time $t_{cpu}^{lim}$ requirement by leveraging penalty based homotopy method for solving mixed-integer optimization problems online. All three MPC formulations use \eqref{eq:plantmodel} as their prediction model. We use \texttt{CasADi} version 3.6.4 to implement each MPC strategy \cite{Andersson2019casadi}. 

\subsection{Mixed-Integer MPC}
\label{subsec:MI-MPC}
In the standard MI-MPC approach, the optimization problem does not consider uncertainty w.r.t $p(t)$ in \eqref{eq:plantmodel}, instead predicting state evolution given the nominal input matrix $B_0$. The controller is called every $t \in \{0,\tau_c,2\tau_c,\ldots\}$ time steps, solving the mixed integer quadratic program (MIQP),

\begin{subequations}  \label{eq:MI-MPC}
\vspace{-1em}
\begin{align}
&\min_{u_{k}} \sum^{N}_{k=0}  \|y_{k} - y_{k}^r\|_Q^2 + \|u_{k}\|_R^2 \label{eq:MI-MPCobjective} \\
&\quad \text{s.t.}  \nonumber \\
&x_{k+1} = Ax_{k} + B_0u_{k-d},  \quad k \in \{0,\ldots,N\mathcal{-}1\},  \label{eq:MI-MPCxpred} \\
&y_{k} = Cx_{k} ,  \quad k \in \{0,\ldots,N\},  \label{eq:MI-MPCypred} \\ 
&g(y_{k}) \leq 0 ,  \quad k \in \{0,\ldots,N\},  \label{eq:MI-MPCstatecons}\\
&u_{k} \in \{0,1\} ,  \quad k \in \{0, \tau_c, \ldots, (N_c\mathcal{-}1) \tau_c \},\label{eq:MI-MPCinputcons} \\
&u_{k} = 0, \quad k \in \{0, \ldots, N\}  \setminus \{0, \tau_c, \ldots, (N_c - 1)\tau_c\},\label{eq:MI-MPCinputdwellcons} \\
&x_{0} = C^\dagger y(t), \label{eq:MI-MPCinitstate}
\end{align}  
\end{subequations}

\noindent wherein we account for the multi-rate $\tau_c  = 100\tau_s \in \mathbb{Z}^+$ by defining control horizon $N_c = N\tau_s/\tau_c  \in \mathbb{Z}^+$ from the prediction horizon $N \in \mathbb{Z}^+$. We solve for $N_c$ discrete inputs $u_{k} \in \{0,1\}$ for $k \in \{0, \tau_c,\ldots , (N_c\mathcal{-}1)\tau_c\}$. Input delay $d$ is incorporated as $u_{k-d}$ in \eqref{eq:MI-MPCxpred}, where $u_{k-d}$ at $k-d<0$ are stored from the previous MPC solution at $t-\tau_c$. Between control decisions $k\notin \{0, \tau_c,\ldots , (N_c\mathcal{-}1)\tau_c\}$, $u_k = 0$. The input $u(t) = u_0$ is then applied to the system. This multi-rate scheme with input delay applies for all three MPC strategies, where $d = 135$. We take $N = 500$ and $N_c =5$ to be long enough to ensure practical stability without the use of a specifically designed terminal cost \cite{maciejowski2002}. The state prediction $x_0$ is initialized with $C^\dagger y(t)$ where $y(t) = n_{e}(t,\rho)$ and $C$ is the output matrix in \eqref{eq:plantoutput}. The constraint \eqref{eq:MI-MPCstatecons} enforces $n_{e,edge} \leq n_{e,lim}$. For all three MPC strategies, the penalty matrix $Q = \text{diag}(10\cdot\mathbf{1}_{40},10^{-4}\cdot\mathbf{1}_{nx-40})$ to yield reference tracking within the region $\rho\in [0,0.4]$, and $R = 1$ to reduce the number of pellets fired (this tuning is the same for \eqref{eq:msMI-MPCobjective} and \eqref{eq:pthmpcobjective}). This certainty-equivalent MI-MPC formulation, however, risks violating the $n_{e,lim}$ constraint \eqref{eq:MI-MPCstatecons} as it does not consider uncertainty in the pellet deposition profile.

\begin{figure}[t!]
    \vspace{2mm}
    \centering
    \includegraphics[width=\columnwidth]{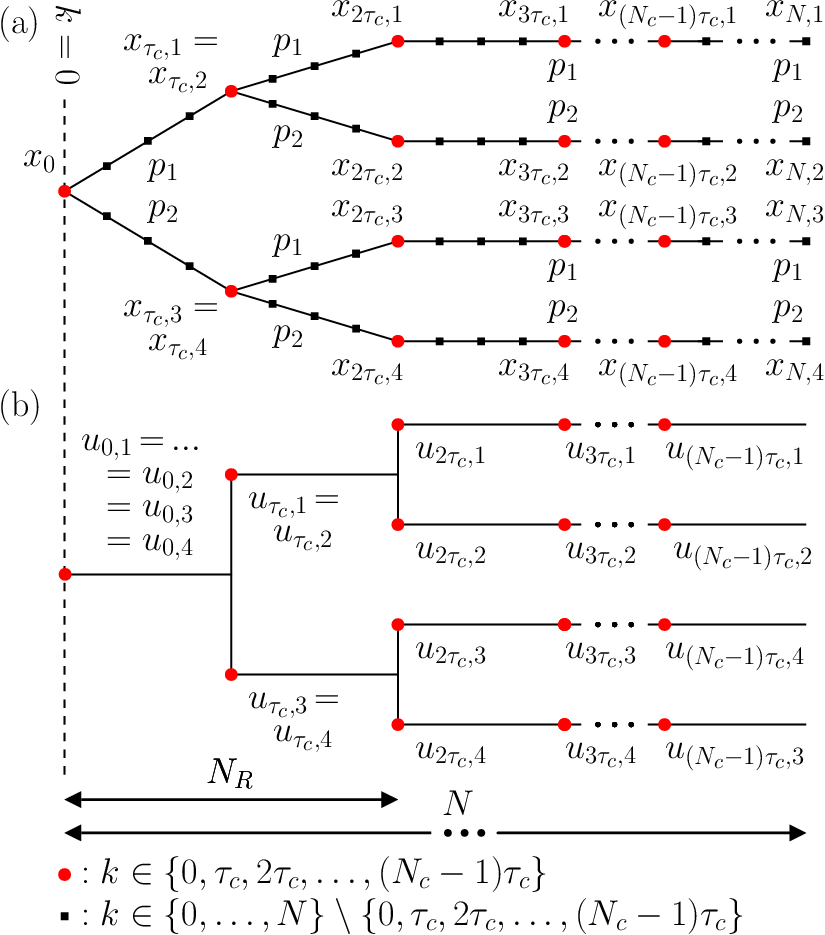}
    \caption{Example scenario tree for msMI-MPC with $N_R = 2$, $n_p=2$, and $\tau_c = 5\tau_s$, branching at control nodes (red). }
    \label{fig:msmpc}
    \vspace{-0.5cm}
\end{figure}

\subsection{Multi-Stage Scenario MI-MPC}
\label{subsec:MSMPC}
The multi-stage scenario MI-MPC (msMI-MPC) builds on the MI-MPC to \textit{explicitly} account for the uncertainty w.r.t where the pellet deposits its particles in the plasma through the linearly parametrized input matrix $B(p_j) =p_j + B_0 \in \mathbb{R}^{n_x \times n_u}$, with $p_j \in \mathbf{P}$. The parameter uncertainty is represented by a scenario tree as depicted in Fig. \ref{fig:msmpc}. Each node signifies a prediction step, and the branches represent parametric uncertainty as extreme disturbance realizations that the system might encounter in future states. A scenario is defined as the entire path from $x_0$ to $x_{N,j}$, where $j \in \{1,\ldots,S\}$ indicates the scenario identifier for $S$ scenarios. For a control horizon of $N_c$, $n_p$ parameter realizations lead to an exponential branch expansion of $S =n_p^{N_c}$, and therefore ${n_u}^{(n_p^{N_c})}$ decision variables. Thus, it is common to assume a constant parameter value after a certain number of predictions $N_R$, known as the \textit{robust horizon} of the controller \cite{lucia2013}. As the parameter realizations are input affine, the scenario tree only branches at control nodes up to $k/\tau_c < N_R-1$, limiting the problem to $S = n_p^{N_R}$ scenarios.

Fig. \ref{fig:msmpc} illustrates an example scenario tree for $n_p = 2$, $N_R=2$. At each decision node, where the future parameter realization is uncertain, (for all $u_{k,j}$ where $k \in \{0, \tau_c, \ldots ,(N_R-1)\tau_c \}$), the control actions must be identical. This ensures uniformity among decisions that branch from a shared node (red in Fig. \ref{fig:msmpc}) within the robust horizon. Consequently, the first control input, $\tilde{u}_0$, remains the same across all scenarios and is the one implemented. This is formally reflected through the \textit{non-anticipativity} constraint \eqref{eq:msMI-MPCnonantcons}, where the matrix $E_k$ equates all initial inputs for a given initial state, encoding causality. For example, in Fig. \ref{fig:msmpc}b, at prediction step $k=\tau_c$, $E_{\tau_c} = [1~{-}1~0~0;~0~0~1~{-}1]$ encodes $u_{\tau_c,1} = u_{\tau_c,2}$ and $u_{\tau_c,3} = u_{\tau_c,4}$, though $u_{\tau_c,1}$ need not equal e.g. $u_{\tau_c,4}$.

At each time step $t$, the controller solves the MIQP \eqref{eq:msMI-MPC} below, applying $u(t) = \tilde{u}_0$, with $u_{0,j} = \tilde{u}_0 \; \forall j$. Here, $w_j$ is the weight given to each scenario. In this work, we use a uniform weighting of $w_j = \frac{1}{S}$ and $N_R=1$. Similar to noted implementations the of multi-stage scenario MPC \cite{scokaert1998scenarioMPC, lucia2013, Mayne2016}, we choose $N$ which is large enough to ensure constraint satisfaction without the use of invariant terminal sets \cite{grune2011stability}.

\begin{subequations}\label{eq:msMI-MPC}
\vspace{0.5em}
\begin{align}
&\min_{u_{k,j}} \sum_{j=1}^S w_j \sum^{N}_{k=0} \|y_{k,j} - y_{k}^r\|_Q^2 + \|u_{k,j}\|_R^2 ~ \cdots \label{eq:msMI-MPCobjective} \\
&\quad \text{s.t.}  \nonumber \\
&x_{k+1,j} = Ax_{k,j} + B(p_j)u_{k-d,j}, ~ k \in \{0,\ldots,N\mathcal{-}1\}, \label{eq:msMI-MPCxpred} \\
&y_{k,j} = Cx_{k,j}, \quad k \in \{0,\ldots,N\}, \label{eq:msMI-MPCypred} \\ 
&\mathbf{g}(y_{k,j}) \leq 0, \quad k \in \{0,\ldots,N\}, \label{eq:msMI-MPCstatecons} \\
&u_{k,j} \in \{0,1\}, \quad k \in \{0, \tau_c, \ldots, (N_c\mathcal{-}1) \tau_c \},\label{eq:msMI-MPCinputcons} \\
&u_{k,j} = 0, ~~ k \in \{0, \ldots, N\}  \setminus \{0, \tau_c, \ldots, (N_c - 1)\tau_c\},\label{eq:msmpcinputdwellcons} \\
&\sum_{j=1}^S E_ku_{k,j} = 0, \quad k \in \{0, \tau_c, \ldots, (N_c\mathcal{-}1) \tau_c \},  \label{eq:msMI-MPCnonantcons} \\
&x_{0,j} = C^\dagger y(t), \label{eq:msMI-MPCinitstate} \\
& \quad j \in \{1,\ldots,S\}. \nonumber 
\end{align}  
\end{subequations}

\begin{figure}[b!]
\vspace{0em}
\centering
\parbox{\linewidth}{\includegraphics[width=\linewidth]{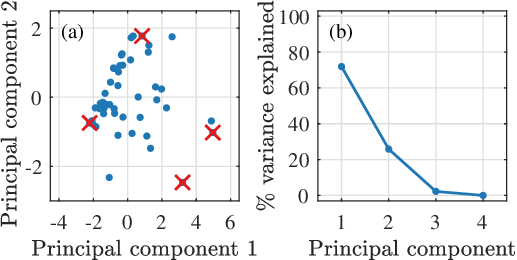}}
\caption{Data-driven parameterization of $p_j$. (a) Score plot using the first two principal components of each data point (blue) computed with PCA. The red x's indicate the chosen scenario realizations. (b) The $\%$ of the total variance explained by each principal component.}
\label{fig:pca}
    \vspace{-0.2cm}
\end{figure}

Given that $p(t) \in \mathbb{R}^4$ in \eqref{eq:plantmodel}, considering all combinations of extreme values across each component results in 16 possible parameter realizations,  which is computationally prohibitive for real-time control applications.

Instead, we construct a reduced set of parameter realizations that are representative of the variation in $\mathbf{P}$ using the data-driven scenario selection algorithm proposed in \cite{krishnamoorthy2018data}. By accounting for correlations between the parameter realizations, principal component analysis (PCA) identifies the orthogonal directions that represent maximal variance in the data. In contrast, element wise minimum or maximum operations treat components independently and thus results in unnecessarily conservative disturbance scenarios. Using PCA, we thus construct physically meaningful (extreme-case) realizations of the disturbances.
% Instead, it is desirable to consider only the realizations of $p_j \in \mathbf{P}$ needed to handle uncertainty. We utilize the data-driven scenario selection algorithm proposed in \cite{krishnamoorthy2018data} to choose parameter realizations. 
% The idea is to perform a PCA operation on the parameter matrix $\mathbf{P}$ to exploit the correlation structure that exists between the parameter components. By projecting parameter space on to its principal components, we are able to identify the most informative parameter realizations that explain a majority of the variance in the parameter set. 

As shown in Fig \ref{fig:pca}b, the first two principal components explain over $97.3\%$ of the total variance in $\mathbf{P}$. Fig. \ref{fig:pca}a plots the projection of different elements in $\mathbf{P}$ onto the first two principal components. We construct the scenario set $S$ by choosing the parameter realizations that are farthest from mean along either principal component, thereby reducing the number of scenarios from 16 to 4 while preserving the most significant variations. 

%Fig. \ref{fig:pca}a presents the score plot of the first two principal components of $\mathbf{P}$ in the principal component space. As shown in Fig \ref{fig:pca}b, the first two principal components explain over $97.3\%$ of the total variance in $\mathbb{P}$ (Fig. \ref{fig:pca}b). The scores with values furthest from the mean along the first two principal components (red x's in \ref{fig:pca}a) each correspond to a parameter realizations in $\mathbf{P}$. By choosing only these parameter realizations $p_j \in \{p_1,p_2,p_3,p_4\}$, we reduce the number of scenarios from 16 to 4 while preserving the most significant variations. 

\begin{table}[t!]\label{alg:PTH}
\vspace{1em}
\centering
\setlength\tabcolsep{0pt}
\begin{tabular*}{0.65\linewidth}{rl} 
\toprule[2pt]
\multicolumn{2}{l}{\textbf{Algorithm 1} PTH-MPC} \\ \midrule
1: & \hspace{1mm} \textbf{set} $i\leftarrow0$, $\gamma_i \leftarrow 0$, and $\beta_i \leftarrow 0$\\
2: & \hspace{1mm} \textbf{while} $\left( u_{k,j} > \epsilon ~ \textrm{\textbf{or}} ~ 1 - u_{k,j} > \epsilon \right) \ldots$ \\
& \hspace{5mm} $\forall k \in \{0,\ldots,N-1\}, ~ \forall j \in \{1, \ldots, S\},$ \\
& \hspace{5mm} \textbf{and} $i < i_{\textrm{{max}}}$ \textbf{do:} \\
3: & \hspace{3mm} \textbf{solve} the optimization problem \eqref{eq:pthmpc}\\
4: & \hspace{3mm} \textbf{set} $i \leftarrow i+1$ \\
5: & \hspace{3mm} \textbf{set} $\beta_i \leftarrow \beta_{\text{init}} \beta_{\text{inc}}^{i-1}$, $\gamma_i \leftarrow \gamma_{\text{init}} \gamma_{\text{inc}}^{i-1}$   \\
6: & \hspace{1mm} \textbf{end while} \\
7: & \hspace{1mm} \textbf{if} $i=i_{\text{{max}}}$ \textbf{set} $u_{k,j} \leftarrow 0$\\
\bottomrule
\multicolumn{2}{p{0.6\linewidth}}{$\epsilon$ is sufficiently small such that the rounding error on $u_{k,j}$ is negligible.}\\
\end{tabular*}
\vspace{-1mm}
\end{table}

\subsection{Multi-Stage Scenario PTH-MPC}
\label{subsec:RMSMPC}
Even after minimizing $S$ using data-driven parameter selection, the added computational complexity of msMI-MPC remains too slow for real time application in fusion devices where control decisions need to be made every $100$ ms. We now reduce the computational complexity of the mixed-integer optimization problem itself using the modified penalty-term homotopy algorithm for MPC (PTH-MPC). The PTH algorithm, proposed in \cite{sager2006} for \textit{unconstrained} MIPs and modified in \cite{orrico2023} for \textit{constrained} MPC applications (c.f. Algorithm 1), relaxes the integrality constraint $u_{k,j} \in \{0,1\}$ in \eqref{eq:msMI-MPCinputcons} to $u_{k,j} \in [0,1]$ and adds a complimentarity penalty term in the cost function \eqref{eq:pthmpcobjective}, allowing us to solve a series of quadratic programs (QPs) instead of an MIQP, as outlined in \cite{orrico2023}. The resulting multi-stage scenario PTH-MPC (msPTH-MPC) is significantly less computationally expensive to solve than msMI-MPC. For msPTH-MPC, \eqref{eq:msMI-MPC} is modified for Algorithm 1 as,

\begin{subequations}\label{eq:pthmpc}
\vspace{-1em}
\begin{align}
&\min_{u_{k,j}} \sum_{j=1}^S w_j \sum^{N}_{k=0} \|y_{k,j} - y_{k}^r\|_Q^2 + \|u_{k,j}\|_R^2 ~ \cdots \label{eq:pthmpcobjective} \\
 &\quad \cdots  + ~  u_{k,j}\beta_i (1-u_{k,j})^{\top} - \gamma_i \ln \left(-g(y_{k,j}) \right) ,\nonumber \\ 
%J(y_{k,j}, u_{k,j} )\label{eq:msPTH-MPCobjective} \\
&\quad \text{s.t.}  \nonumber \\
&x_{k+1,j} = Ax_{k,j} + B(p_j)u_{k-d,j}, ~ k \in \{0,\ldots,N\mathcal{-}1\}, \label{eq:pthmpcxpred} \\
&y_{k,j} = Cx_{k,j}, \quad k \in \{0,\ldots,N\}, \label{eq:pthmpcypred} \\ 
&\mathbf{g}(y_{k,j}) \leq 0, \quad k \in \{0,\ldots,N\}, \label{eq:pthmpcstatecons} \\
&u_{k,j} \in [0,1], \quad k \in \{0, \tau_c, \ldots, (N_c\mathcal{-}1) \tau_c \},\label{eq:pthmpcinputcons} \\
&u_{k,j} = 0, ~~ k \in \{0, \ldots, N\}  \setminus \{0, \tau_c, \ldots, (N_c - 1)\tau_c\},\label{eq:pthmpcinputdwellcons} \\
&\sum_{j=1}^S E_ju_{k,j} = 0, \quad k \in \{0, \tau_c, \ldots, (N_c\mathcal{-}1) \tau_c \},  \label{eq:nonantcons} \\
&x_{0,j} = C^\dagger y(t), \label{eq:pthmpcinitstate} \\
& \quad j \in {1,\ldots,S}. \nonumber 
\vspace{-1em}
\end{align}  
\end{subequations}

In Algorithm 1, \eqref{eq:pthmpc} is solved successively by increasing the penalty term $u_{k,j}\beta_i(1-u_{k,j})$ in \eqref{eq:pthmpcobjective}. To ensure $u_{k,j}$ converges to $\{0,1\}$ when $y_{k,j}$ nears constraints, the modified PTH algorithm includes a logarithmic penalty term $ - \gamma_i \ln \left(-g(y_{k,j}) \right)$. In Algorithm 1, we take $\gamma_\text{init} = 32$, $\gamma_\text{inc} = 2$, $\beta_\text{init} = 32$, $\beta_\text{inc} = 2$, tuned heuristically as described in \cite{orrico2023} (see for further details on the modified PTH algorithm). 

\begin{figure}[t!]

\vspace{0.5em}
\centering
\parbox{\linewidth}{\includegraphics[width=0.94\columnwidth]{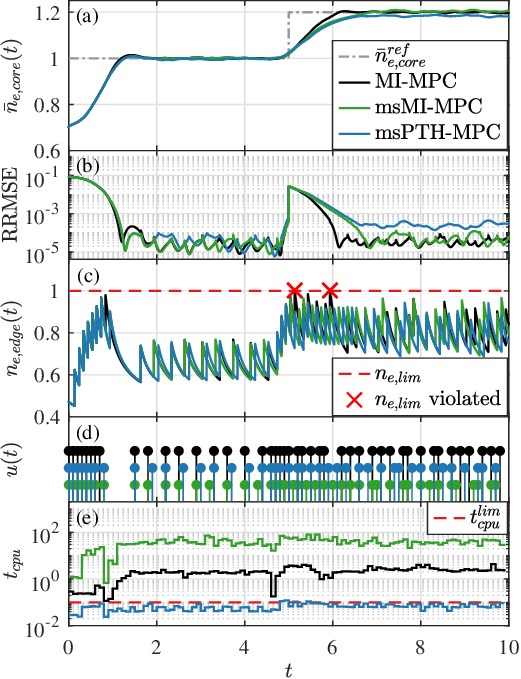}}
\caption{Simulation results with MI-MPC (black), msPTH-MPC (blue), and msMI-MPC (green). (a) $\bar{n}_{e,core}^{ref}$ compared to $\bar{n}_{e,core}(t)$  $[10^{20}$ m$^{-3}]$. (b) The RRMSE of $n_e(t,\rho)$, averaged over $\rho = [0,0.4]$. (c) The constraint satisfaction at $n_{e,edge}(t)$ $[10^{20}$ m$^{-3}]$ w.r.t $n_{e,lim}$ (red). MI-MPC violations of $n_{e,lim}$ are denoted with red x's (note that consecutive violations overlap). (d) Control decisions for each controller. (e) $t_{cpu}$ [s] required to compute each MPC strategy, compared to $t_{cpu}^{lim}$ (red). }
\label{fig:simulation}
\vspace{-0.5cm}
\end{figure} 

\begin{figure}[t!]
\vspace{0.5em}
\centering
\parbox{\linewidth}{\includegraphics[width=0.94\columnwidth]{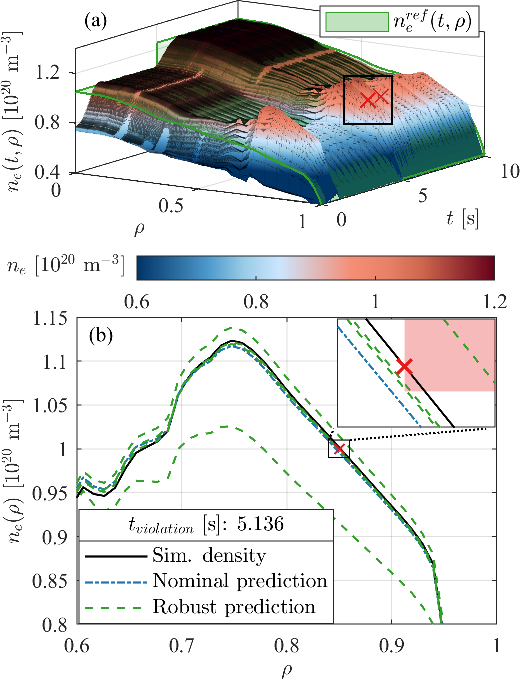}}
\caption{$n_e(t,\rho)$ over the simulation with MI-MPC. Jumps in $n_e(t,\rho)$ occurring near $\rho = 0.8$ each time a pellet is fired exhibits the variation in the ablation profile varies captured in $p(t)$. Each $n_{e,lim}$ violation is labeled as a red x. The black frame region corresponds to (b). (b) The plant $n_e(t,\rho)$ (black) compared to the $n_e(t,\rho)$ predicted by MI-MPC at $t = 5.136$ s. The zoomed frame shows the true $n_e(t,\rho)$ exceeding $n_{e,lim}$, while the MI-MPC-predicted $n_e(t,\rho)$ does not. The msMI-MPC scenarios predicted from the same $y(t)$ and decision $u(t)$ as MI-MPC (green) \textit{do} predict that $n_e(t,\rho)$ will violate $n_{e,lim}$. }
\label{fig:violation}
\vspace{-0.5cm}
\end{figure}

\section{Results}
\label{sec:results}

For each MPC strategy, the density controller must track two steps in $n_e^{ref}(t,\rho)$ over a $10$ s simulation, with $\bar{n}_{e,core}(t) = 1 \times 10^{20}$ m$^{-3}$ from $t = 0\rightarrow 5$ s and $\bar{n}_{e,core}(t) = 1.2 \times 10^{20}$ m$^{-3}$ from $t = 5\rightarrow 10$ s. The plant simulator is \eqref{eq:plantmodel} with random realizations of $p(t) \in_R \mathbf{P}$. The controller must minimize relative root mean square error (RRMSE) of $n_e(t,\rho)$ w.r.t $n_e^{ref}(t,\rho)$ over $\rho = [0,0.4]$ without violating $n_{e,lim}$ at $n_{e,edge}(t)$ or exceeding the maximum $t_{cpu}^{lim}=100$ ms. These computations were performed on an Intel(R) Core\textsuperscript{TM} i5 processor running at 2.40 GHz with 15.7 GB of usable RAM. Each strategy is warm started with its respective solution at the previous timestep. The MI-MPC and msMI-MPC strategies are solved using the \texttt{BONMIN} solver \cite{bonami2008} and msPTH-MPC is solved using \texttt{qpOASES} \cite{Ferreau2014}.

Fig. \ref{fig:simulation}a and \ref{fig:simulation}b give the reference tracking performance of each strategy. The mean RRMSE over the simulation of the nominal MI-MPC is $0.549\%$, compared to $0.582\%$ for msPTH-MPC and $0.569\%$ for msMI-MPC. While MI-MPC exhibits the least tracking error, this comes at the cost of MI-MPC violating $n_{e,lim}$ 4 times over the simulation (at $t = \{ 5.136,5.137,5.936,5.937\}$ s), as shown in Fig. \ref{fig:simulation}c. In contrast, the msMPC methods do not violate $n_{e,lim}$ as they ensure constraint satisfaction over each scenario.

We examine how the scenario realizations account for parametric uncertainty in Fig. \ref{fig:violation}. In Fig. \ref{fig:violation}a, we show the evolution of $n_e(t,\rho)$ over the simulation for the nominal MI-MPC controller. We take the profile at the first violation of $n_{e,lim}$ and plot the profile region near the plasma edge in Fig. \ref{fig:violation}b. For the plant $n_{e,edge}(t)$ exceeds $n_{e,lim}$ at $t_{violation} = 5.136$ s, as shown in the zoomed portion of Fig. \ref{fig:violation}. The pellet was fired by the nominal MI-MPC controller because the nominal predicted $n_{e,edge}(t_{violation}) < n_{e,lim}$. However, given the same initial $y(t)$ as the MI-MPC controller at the control time step when the pellet was fired, 3/4 of the scenarios of \eqref{eq:msMI-MPCxpred}-\eqref{eq:msMI-MPCypred} and \eqref{eq:pthmpcxpred}-\eqref{eq:pthmpcypred} predict $n_{e,edge}$ exceeding $n_{e,lim}$ in Fig \ref{fig:violation}b. Consequently, both msPTH-MPC and msMI-MPC would \textit{not} have fired a pellet given the same initial condition as the nominal MI-MPC controller, and therefore would not have violated $n_{e,lim}$.

Lastly, we compare computational costs in Fig. \ref{fig:simulation}e. Both the MI-MPC and msMI-MPC strategies are very computationally expensive, with $t_{cpu}^{max} = 4.346$ s and $t_{cpu}^{max} = 80.54$ s, respectively. In contrast, $t_{cpu}^{max} = 0.121$ s for the msPTH-MPC, very nearly below the $t_{cpu}^{lim}$ of 0.1 s. With a mean $\bar{t}_{cpu}=0.0646$ s for over the simulation, the msPTH-MPC could meet the $t_{cpu}^{lim}$ for real-time control given improved computing hardware and/or specialized QP solvers. 

\section{Conclusion}
\label{sec:conclusion}

We compared three mixed-integer MPC (MI-MPC) strategies for the case study on density profile control in the ITER tokamak, a nonlinear PDE system with safety-critical constraints, input delays, and discrete actuators under parametric uncertainty. We utilize a multi-stage scenario approach to MPC (msMPC) to handle actuator uncertainty. We limit computational complexity by using dynamic mode decomposition with control on system identification data to produce a reduced order model for control and applied a principal component analysis-based method to manage the added complexity of the msMI-MPC scenario tree. We further improve real-time viability by combining msMPC with the modified penalty term homotopy algorithm for MPC (PTH-MPC). In simulations with MI-MPC, msMI-MPC, and msPTH-MPC, only the msPTH-MPC strategy achieved good reference tracking without violating density constraints while staying near the computational time limit. Hence, msPTH-MPC offers a promising strategy for predictive density control in ITER. As this work is limited to linear plant simulation of a true tokamak plasma, efficacy of the msPTH-MPC must be evaluated in JINTRAC simulations (wherein the nonlinear particle transport drives the structural uncertainty \cite{orrico2025}).

\bibliography{CDC_2025}
\bibliographystyle{IEEEtran}

\end{document}